# Parametrically Excited Time-varying Metasurfaces for Second Harmonic Generation

Xuexue Guo and Xingjie Ni[*]

Department of Electrical Engineering, the Pennsylvania State University, University Park, PA 16802

[*]xingjie@psu.edu

Parametric oscillation is a fundamental concept that underlies nonlinear wave-matter interactions, leading to generation or amplification of new frequency components [1]. Using a temporal modulation generated by the heterodyne interference of a control direct current (dc) electric field and an optical field, we resonantly drive large-amplitude oscillations of amorphous silicon meta-atoms – nanoscale building blocks of an optical metasurface. We observed gigantically enhanced second harmonic generation (SHG) with an electric-field-controlled on/off ratio over $10^4$ and an ultra-high modulation depth of 40500%$V^{-1}$. A simple Mathieu equation involving a time-dependent resonance captures most features of experiment data (for example, SHG intensity has super-quadratic dc field dependence), and provides insight into SHG efficiency enhancement through parametric modulations. Our work provides a compact, electric-tunable and CMOS (Complementary Metal-Oxide Semiconductor) compatible approach to boosting and dynamically controlling nonlinear light generations on a chip. It holds great potential for applications in optical communications and signal processing.

**Introduction**

Parametric oscillation – originated from the time varying property of a system – has been an interesting topic explored in many research areas ranging from mechanical engineering[1–3], quantum physics[4], plasma physics[5] to electronics[6] and radio-frequency engineering[7]. In particular, it lays the basis for constructing optical parametric oscillator (OPO) and amplifier (OPA), which are key to every aspects of optical and photonic applications[8,9]. However, current systems[10] are limited by strict phase matching conditions, bulk nonlinear crystals and large optical resonators. *It remains a challenge to achieve strong and tunable OPOs on a compact platform compatible to CMOS technology*. Despite great effort in miniaturization and integration of OPO/OPA using silicon photonics technology[11,12], the devices are inherently constrained by phase matching conditions and require millimeter interaction lengths. Recent studies on nonlinear metasurfaces - consist of artificial subwavelength building blocks known as meta-atoms - have shown great potential for increasing nonlinear generation without phase matching constraint[13–16], and agile control over the properties of nonlinear signals[17]. Leveraging the resonance enhanced local field, nonlinear signals are boosted within a subwavelength interaction length, which is essential for reducing device footprint and integration. However, *the resonant property of meta-atoms is in general fixed and difficult to tune after fabrication, thus impeding the realization of efficient OPO using metasurfaces*.

Here, *a prominent time-varying resonance is generated by optimizing the dynamic phase change ($\Delta\varphi$) of amorphous silicon (α-Si) meta-atoms under parametric excitation, thus creating strong parametric oscillation and tremendously enhanced second harmonic generation (SHG)*. The parametric excitation is generated by the travelling-wave interference of a control dc electric field and an optical pump field. Taking advantage of the large nonlinear Kerr effect of α-Si meta-atoms, an ultrafast (196 THz) temporal phase modulation ($\Delta\varphi \sim 0.7$) was achieved with a subwavelength interaction length (~350nm). This leads to a dc electric field controlled on/off ratio over $10^4$ and a modulation depth of 40500%V$^{-1}$, which are two to three orders of magnitude greater than those of conventional electric field induced second harmonic (EFISH) demonstrations[18–20]. Using a forced Mathieu equation[1], we predicted that the SHG intensity scales with $J_1^2(\Delta\varphi) / J_0^2(\Delta\varphi)$ and has a super-quadratic dc electric field dependence, which agrees well with our experiment observations. Leveraging the time-varying phase/resonance frequency modulation of meta-atoms, our metasurfaces represent a new paradigm for electrically controlled efficient nonlinear light generation. It holds great promise for constructing ultra-compact photonic/optoelectronic devices for on-chip coherent light generation, ultrafast optical switching, and information processing.

**Results**

**Theory and Design.** As illustrated in Fig. 1, the parametric oscillation at frequency $\Omega$ is generated by a parametric excitation in the material, where each atom experiences a tiny wiggling of the phase $\Delta\varphi\cos(\Omega t)$. An input wave ($A\mathrm{e}^{-i\omega t}$) interacting with the material will acquire a time dependent phase expressed as $A\mathrm{e}^{-(i\omega t + i\Delta\varphi \cos(\Omega t))}$. The parametric oscillation in a natural material (Fig. 1a) is mostly non-resonant, resulting in small $\Delta\varphi$. Using Taylor expansion, the output wave can be decomposed into two sidebands proportional to $(A\Delta\varphi/2)\mathrm{e}^{-i(\omega+/-\Omega)t}$. In contrast, in an artificial material where the constituent meta-atoms respond resonantly to the parametric excitation, a much stronger phase oscillation can be generated (Fig. 1b). As a result, an identity involving Bessel functions (*i.e.* Jacobi-Anger expansion) are used to derive the field amplitude of two sidebands, leading to $J_1(\Delta\varphi)\mathrm{e}^{-i(\omega+/-\Omega)t}$. It worths noting that $J_1(\Delta\varphi)$ determines the generation efficiency of the new frequency components, and can be designed to tune and enhance the nonlinear generation.

To create the parametric excitation in our device (Fig. 1c), a fundamental optical pump field ($\omega$) interferes with a dc electric field applied on the meta-atoms by interdigitated electrodes array, so that $\Omega = \omega$. We used dielectric cuboid nanoantennas made of α-Si as meta-atoms (Fig. 2a), because α-Si has high refractive index, large Kerr nonlinearity, and low optical loss in the telecommunication wavelengths. Thanks to nonlinear Kerr effect, the travelling-wave intensity modulation leads to a time dependent refractive index change of the meta-atoms, and thus induces a dynamic phase shift. The meta-atoms are specially tailored to achieve large $\Delta\varphi$. We simulated the phase shifts ($\varphi_{\mathrm{linear}}$ and $\varphi_{\mathrm{nonlinear}}$) induced by meta-atoms of various sizes ($l_\mathrm{x}$ and $l_\mathrm{y}$) upon incident light ($\lambda = 1550$nm) at low intensity (linear) and at high intensity (0.4 GW/cm$^2$, nonlinear), respectively, and then calculated the dynamic phase shift change ($\Delta\varphi = \varphi_{\mathrm{nonlinear}} - \varphi_{\mathrm{linear}}$). The design of $l_\mathrm{x} = 600$ nm and $l_\mathrm{y} = 650$ nm was chosen to provide the largest $\Delta\varphi$ (Fig. 2b). The simulated $\Delta\varphi$ shows a super-linear dependence on the modulation intensity (Fig. 2c), leading to a large $\Delta\varphi$ at a relatively low modulation intensity. This time-varying phase $\Delta\varphi\cos(\Omega t)$ can be linked to a time-varying resonance shift. To get a quantitative explanation of SHG, we treat the meta-atom as a Lorentz oscillator with a time-dependent resonance frequency, and describe its motion using a forced Mathieu equation (see supplementary information):

$$m\ddot{\tilde{x}} + \frac{m\omega_0}{Q}\dot{\tilde{x}} + m\left(\omega_0^2 + \alpha\cos(\omega t)\right)\tilde{x} = \tilde{F}(t), \tag{1}$$

where $m$ is the effective mass and $Q$ is the quality factor of the meta-atom, and $\tilde{F}(t) = F_0 e^{-j\omega t} + c.c.$ is the external driving force imposed by the third optical pump field. The temporal modulation induced by dc and optical pump field is expressed as $\alpha\cos(\omega t) = \frac{\omega_0^2 J_1(\Delta\varphi)}{Q J_0(\Delta\varphi)}\cos(\omega t)$, where $\alpha$ is a coefficient related to the resonance frequency modulation depth, and $J_n(\Delta\varphi)$ is the Bessel function of order $n$. By solving equation (1) at $\omega_0 = \omega$, we obtain the nonlinear polarization at $2\omega$

$$P^{(2)} = Z x^{(2)}(2\omega) \approx \frac{Z F_0 J_1(\Delta\varphi)}{8\sqrt{2} m\omega_0^2 J_0(\Delta\varphi)} e^{-j2\omega t}, \tag{2}$$

where $Z$ is the effective dipolar charge. It is evident that SHG is determined by the time-varying phase modulation through $\frac{J_1^2(\Delta\varphi)}{J_0^2(\Delta\varphi)}$, which can be optimized in our design to boost the SHG efficiency. In conventional EFISH, $\Delta\varphi$ is normally very small and is linearly proportional to dc electric field under weak modulation (Fig. 2c), therefore, EFISH intensity has a quadratic dc field dependence [20].

**Parametrically enhanced SHG.** The linear response of the metasurface was characterized by measuring its reflectance spectra with and without a dc electric field, respectively (Methods). The reflectance spectra show a prominent resonance around 1540 nm (Fig. 3a). Evidently, dc electric field does not shift the reflectance spectra, indicating that dc field alone cannot affect the resonant properties of the meta-atom.

To create the parametric oscillation, a femtosecond pulsed laser was focused onto the meta-atoms array biased by a dc electric field. As the dc control voltage $V_c$ increased, the SHG signal increased super-quadratically (Fig. 3b). Experimental results agree well with eq. (2) validating that the significantly boosted SHG arises from the enhanced parametric oscillation of meta-atoms. The SHG signal exhibited an on/off ratio of 15000 ( $(I_{SHG}(37V) - I_{SHG}(0V))/I_{SHG}(0V)$ ), which is three orders of magnitude greater compared with other EFISH nanophotonic devices[18–20]. We estimated the effective second order sheet susceptibility $\chi^s_{yyy}{}^{(2)}$ with a z-cut quartz substrate as reference[21–24]

(see supplementary information), and obtained a large $\chi^s{}_{yyy}{}^{(2)}$ around $2.988 \times 10^{-19}$ m$^2$/V, which is comparable to 2D materials[25] with strong nonlinearity. In addition, the electrically tunable SHG has a modulation depth of $I_{SHG}$ (V=37V)/[$\Delta$V$\times I_{SHG}$(V=0V)] ~ 40500%V$^{-1}$, which is the largest among EFISH demonstrations [18,19,26,27] to the best of our knowledge. More interestingly, the 4$^{th}$ order harmonic generation was also observed at a low input intensity of 0.8 GW/cm$^2$ thanks to the strong parametric oscillation (Fig. S3a), even though our device was not tailored for high-order harmonic generation (HHG). Besides, the required intensity is two to three orders of magnitude lower than a recent demonstration on HHG generation on dielectric metasurfaces[28]. Therefore, our time-varying metasurface can be potentially used for enhancing and tuning high-order harmonic generations.

In contrast, SHG with off-resonance modulation (Fig. 3c) showed a much smaller enhancement effect. Due to the absence of a resonantly enhanced local field, $\Delta\varphi$ is a lot smaller, so that $\frac{J_1^2(\Delta\varphi)}{J_0^2(\Delta\varphi)} \approx \left(\frac{1}{2}\Delta\varphi\right)^2$. SHG signal exhibits a quadratic dependence on dc electric field and a low enhancement factor of 270. In another control experiment, meta-atoms that are rotated by 90 degrees ($l_x$ = 650nm and $l_y$ = 600nm) were excited by optical filed polarized perpendicular to the dc electric field. In this situation, a similar local field enhancement effect was induced, however, the effective parametric excitation is weaker because $\chi^{(3)}{}_{yxxy} < \chi^{(3)}{}_{yyyy}$[29]. We obtained a much lower enhancement factor of 50 and a quadratic dc electric field dependence (Fig. 3d). These control experiments further confirm that the enhanced EFISH is attributed to the parametric oscillation created by the dc electric and optical field.

**Electrical switching of SHG.** Fast electrical switching of SHG was also verified by measuring the time-resolved trace of SHG signal modulated by a square-wave voltage signal. We applied short pulses with 5V amplitudes across the IDT electrodes at 1MHz repetition frequency. The SHG signal was detected in a time-resolved measurement, which precisely follows the trace of the voltage pulses (Fig. 4a). The rising and trailing edges were fitted to give a time constant of 5 ns that is limited by the edge time of the function generator. The response time of electrically tunable SHG is theoretically determined by the temporal response nonlinear Kerr effect convoluted with the lifetime (< 10 fs) of the resonant meta-atom. This means our device can operate at a much faster modulation speed.

**Fundamental power dependence and polarization analysis of SHG.** At $V_{dc} = 0$, the measured SHG intensity scaled quadratically with the intensity of the fundamental pump beam. However, as the dc bias was increased to 10V or 30V, it exhibited a sub-quadratic power dependence (Fig. 4b). Even though the effective band gap of α-Si is ~ 1.7 eV, there are exponential band tails with large density of states even below 1.6 eV due to the amorphous nature of silicon[30]. Therefore, a large amount of carrier is generated through enhanced EFISH (~1.6 eV) when biased by a high dc voltage. These free carriers will shield the dc electric field imposed upon the meta-atoms, thus reducing the effective electric field. The shielding effect is stronger at higher dc voltage, which explains the observation that the SHG power dependence decreases as the control dc voltage increases (Fig. 4a). This non-quadratic fundamental power dependence is also universally observed in many EFISH demonstrations on semiconductor materials, in which an interfacial electric field induced by free carrier can modify the built-in electric field of the system [31].

We also studied the polarization properties of the dynamic enhanced EFISH. With dc electric field, the polarization of SHG emission is predominantly along the direction of the dc electric field, whether the input polarization is parallel or perpendicular to the direction of the dc electric field (Fig. 4c and d). When dc electric field is applied, two nonzero effective second order susceptibilities $\chi^{(2)}_{yyy} = \chi^{(3)}_{yyyy} E^y_{dc}$ and $\chi^{(2)}_{yxx} = \chi^{(3)}_{yxxy} E^y_{dc}$ are generated. Regardless of the polarization of the optical field, the EFISH signal is always in y polarization – that is along the dc field direction.

**Discussions**

Despite some similarities to conventional EFISH – in both cases a dc electric field was employed to generate SHG, the mechanism in SHG enhancement is inherently different. As described before, previous demonstrations on EFISH can be categorized into the situation where a small $\Delta\varphi$ was induced in natural materials. This weak parametric oscillation has fundamentally limited the SHG efficiency, as is evident by the electric field controlled on/off ratios of 10 to 100 [18–20] and the quadratic or quasi-linear dependence on control dc field [9]. In contrast, in our design SHG originates from resonantly enhanced parametric oscillation of the phase/resonance of meta-atoms. With a strong temporal phase modulation, an electrically controlled on/off ratio of 15000 and a modulation depth of 40500%V$^{-1}$ were achieved at telecom wavelengths, which are significantly larger than conventional EFISH. In fact, SHG enhancement is only one special case where $\Omega = \omega$.

As long as strong parametric oscillations at $\Omega$ are constructed, nonlinear generations at $\omega +/- \Omega$ can be achieved. In addition, compared with electric field controlled nonlinear generations in 2D materials[27,32,33], our approach not only provides electrical tunability and CMOS compatibility, but also has 3 to 4 orders of magnitude greater efficiency.

We demonstrated parametrically enhanced and electrically tunable SHG based on a time-varying metasurface, which provides a compact, tunable and CMOS compatible platform for tuning and boosting optical nonlinear generations. Our study is among the first to explore and demonstrate the exceptional potential of temporal phase engineering of metasurfaces for nonlinear applications. It paves exciting ways for constructing electrically tunable nonlinear optical devices, such as optical parametric sources, ultrafast optical switches and modulators, and terahertz components, for applications in optical communication, imaging, and sensing, and laser technology.

## Methods

**Device fabrication.** The meta-device exhibiting dynamic phase modulation enhanced SHG was fabricated by the following procedure (Fig. S1). A 100nm deep (70 by 80 µm) trench was patterned by Electron beam lithography (EBL) and inductively coupled plasma - reactive ion etch (ICP-RIE) of the Quartz substrate. Followed by Ebeam evaporation of 100 nm gold into the pre-defined trench and stripping of the photoresist, the back-reflector plate was fabricated. Then, 50 nm of $Al_2O_3$ was deposited by atomic layer deposition (ALD) to prove good electrical insulation. Subsequently, the IDT electrodes and Ebeam markers were created using a sequential process of EBL, ebeam evaporation of 3 nm chromium (adhesion layer) and 50 nm gold, and finally the lift-off process in MICROPOSIT 1165. Then, 350nm of α-Si was deposited by plasma enhanced chemical deposition (PECVD). Nanoantennas were defined by EBL with precise alignment, lift-off of chromium mask, ICP-RIE etch and mask removal in Cr etchant 1020. Finally, the electrical pads were wire bonded to a printed integrated circuit board for connection to external voltage supplier.

**Numerical simulations.** In order to find the optimized parameters of the meta-atom that can create the maximum $\Delta\varphi$ under intensity modulation, we used a commercially available finite element method (FEM) solver package – COMSOL Multiphysics. A periodic boundary condition was applied for a single building block to simplify the model and save computation memory and time. Third-order finite elements and at least 10 mesh steps per wavelength were used to ensure the accuracy of the calculated results. The experimentally obtained optical constants of gold, alumina and α-Si were used to model the back reflector, spacer and the meta-atom. The nonlinear simulation takes two major steps in a customized model: (1) We simulated the spatially dependent permittivity change using an iterative scheme – We first calculated the field distribution in the computational domain with the pump beam ($\lambda$ = 1550nm) incidence, then updated the permittivity in the Kerr medium ($n_2$ = 5 × $10^{-13}$ $cm^2$/W) with the calculated inhomogeneous field, and after that we calculated the field again with modified permittivity. We iterated over the steps above until the change of the field distribution was within a predefined tolerance. (2) We then simulated the structure with the pump-induced permittivity changes with a probe light ($\lambda$ = 1550nm) incidence. We obtained the reflected field distribution of a 1550nm probe beam upon incidence on the nanoantenna whose permittivity has been modified by the pump beam. By varying the sizes ($l_x$ and $l_y$) parameters, we mapped out the phase shift $\varphi$ at different pump intensities (0 and 0.4 GW/$cm^2$), and calculated the phase shift change $\Delta\varphi$ (Fig. 2b).

**Optical measurements**

**Linear measurements.** By replacing the light source in Fig. S2 to a halogen-tungsten lamp and change the dichroic mirror to a non-polarized beam splitter, we performed the linear reflectance measurement on the sample. The broadband light source was coupled into an optical fiber and transformed into a collimated beam. The incident beam was focused through a long working distance, near infrared (NIR) objective (20X, NA = 0.40) from Mitutoyu. The reflected beam was collected by the same objective and reflected by the non-polarized beam splitter towards the spectrometer (Horiba iHR 320) equipped with an InGaAs infrared detector.

**Nonlinear measurements with dc electric field control.** As shown in Fig. S2, an ultrafast pulsed laser (200fs, 80MHz) from an optical parametric oscillator (OPO) seeded by a Ti: Sapphire laser passed through a $\lambda/2$ waveplate to adjust the polarization. Then the laser beam transmitted through a long pass dichroic mirror (cut-on wavelength: 950nm) and then focused onto the sample by a long working distance near infrared (NIR) objective. The reflected SHG signals were collected by the same objective and reflected by the dichroic mirror towards the spectrometer. The electrical signal (dc/ac) was connected to the sample.

In order to estimate the effective second order nonlinear susceptibility ($\chi^{s\,(2)}$) of our device, we used a z-cut quartz crystal substrate as a reference sample. Under the same illumination and collection condition, we measured the SHG from the quartz reference substrate, and calculated $\chi^s_{(2)}$ (supplementary information) by assuming $\chi^{(2)}_{\alpha\text{-quartz}}$ equal to 0.6 pm/V.

**Time-resolved measurement of electrically modulated SHG.** Fig. S4 shows the time - resolved measurement setup. A modulated square wave voltage signal was applied on the sample with a function generator (Angilent 33220). The 1 MHz voltage signal changes between 0V and 5V with a duty cycle of 50%. In order to separate SHG from fundamental optical field and THG, the output signal passed through a transmission grating. An optical fiber was placed in the position for SHG ($\lambda = 765$ nm) collection and then directed the SHG towards a single photon avalanche detector (SPAD). The sync signal of the function generator was used as a start trigger, and the SPAD output was used as a stop in the photon counting module (Picoharp 300, Picoquant). We obtained the accumulated (accumulation time 300s) histogram of the electrical modulated SHG (Fig. 4a).

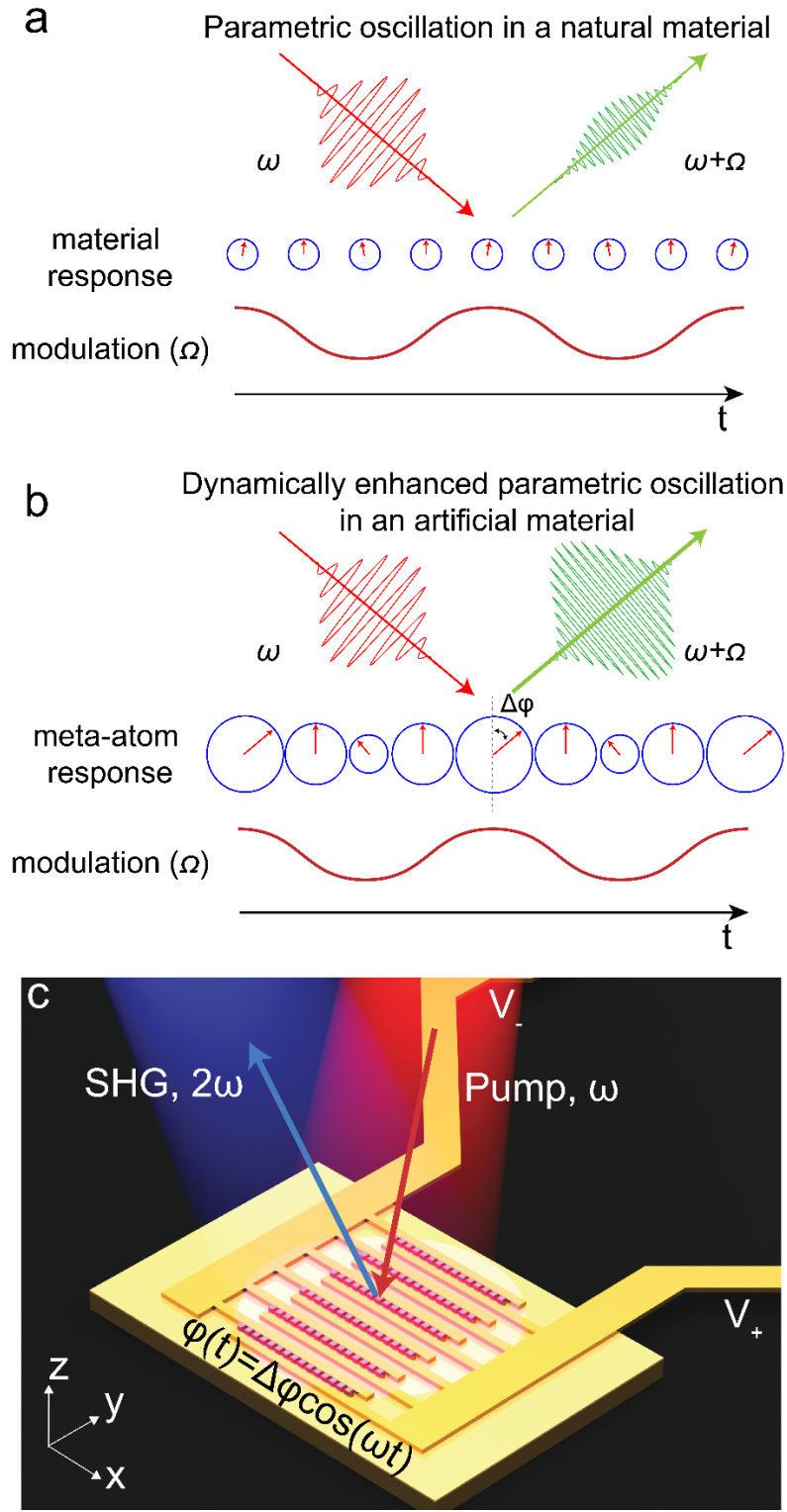

**Fig. 1. Principle of dynamically enhanced parametric oscillation.** (**a**) and (**b**) comparison of the parametric oscillation in a natural material and an artificial material. (**c**) Schematic of the α-Si metasurface biased by IDT electrodes array. The dynamic interference between dc electric field

and fundamental optical field creates a travelling-wave modulation to α-Si meta-atoms, leading to SHG with enhanced efficiency.

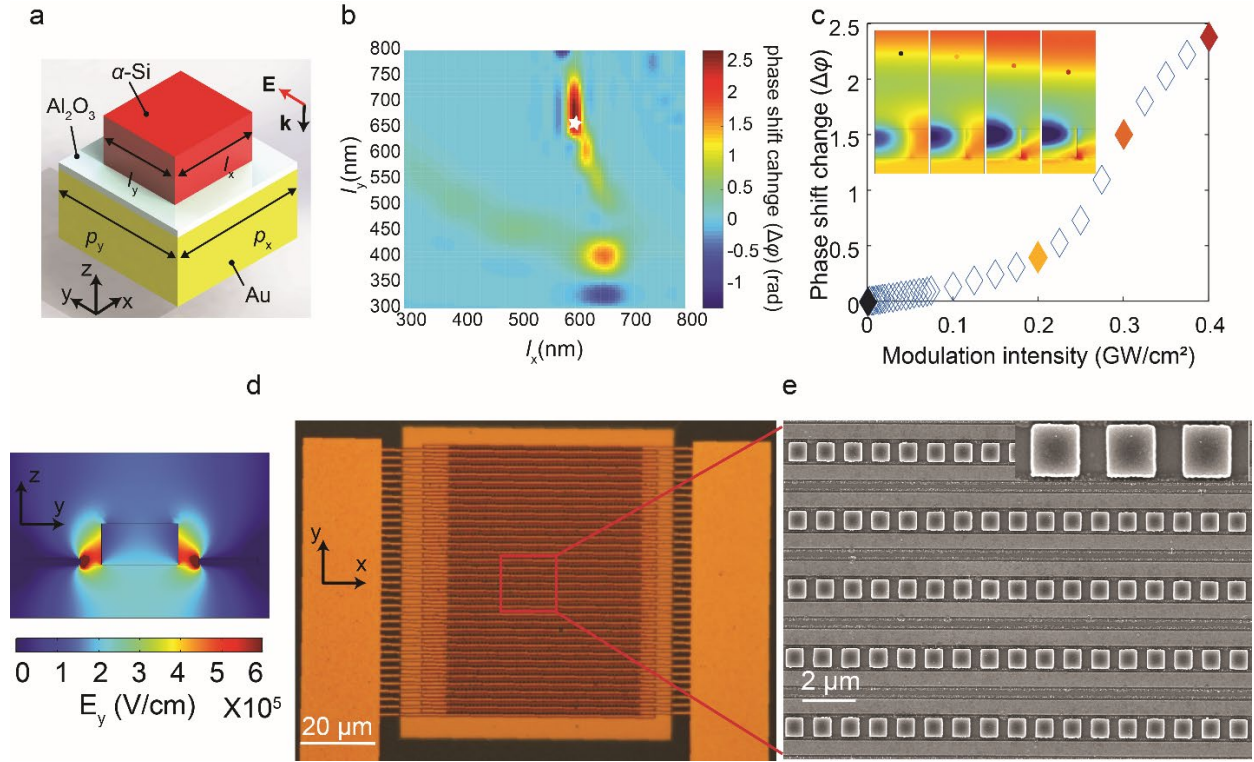

**Fig. 2. Design of the time-varying metasurface**. (**a**) A 3D illustration of a unit cell consists of α-Si nanobar antenna. The thickness of the gold ground plate, the $Al_2O_3$ insulation layer and antenna are 100nm, 50nm and 350nm, respectively. The square lattice periodicity is 1 μm, and the width/length of the antennas is 600 nm/650 nm. (**b**) Simulated dynamic phase shift change ($\Delta\varphi$) of the reflected optical field in a 2D parameter space spanned by $l_x$ and $l_y$. The white diamond marks the size of the designed antenna which exhibits a large $\Delta\varphi$ under the illumination of intense light (0.4 GW/cm²). (**c**) Simulated $\Delta\varphi$ with the change of modulation intensity, which shows a superlinear dependence. The inset displays electric field distribution at different modulation intensity, showing gradual wavefront shifts. (**d**) Simulated dc ($V_{dc}$ = 36 V) electric field distribution ($E_y$) of the IDT electrodes biased nanoantenna. It shows a strong electric field along the edge of the nanoantenna. (**e**) Optical microscope image of the elecrtically controlled metasurface, with a close-up field emission scanning electron microscope (FESEM) image of the nanoantennas array.

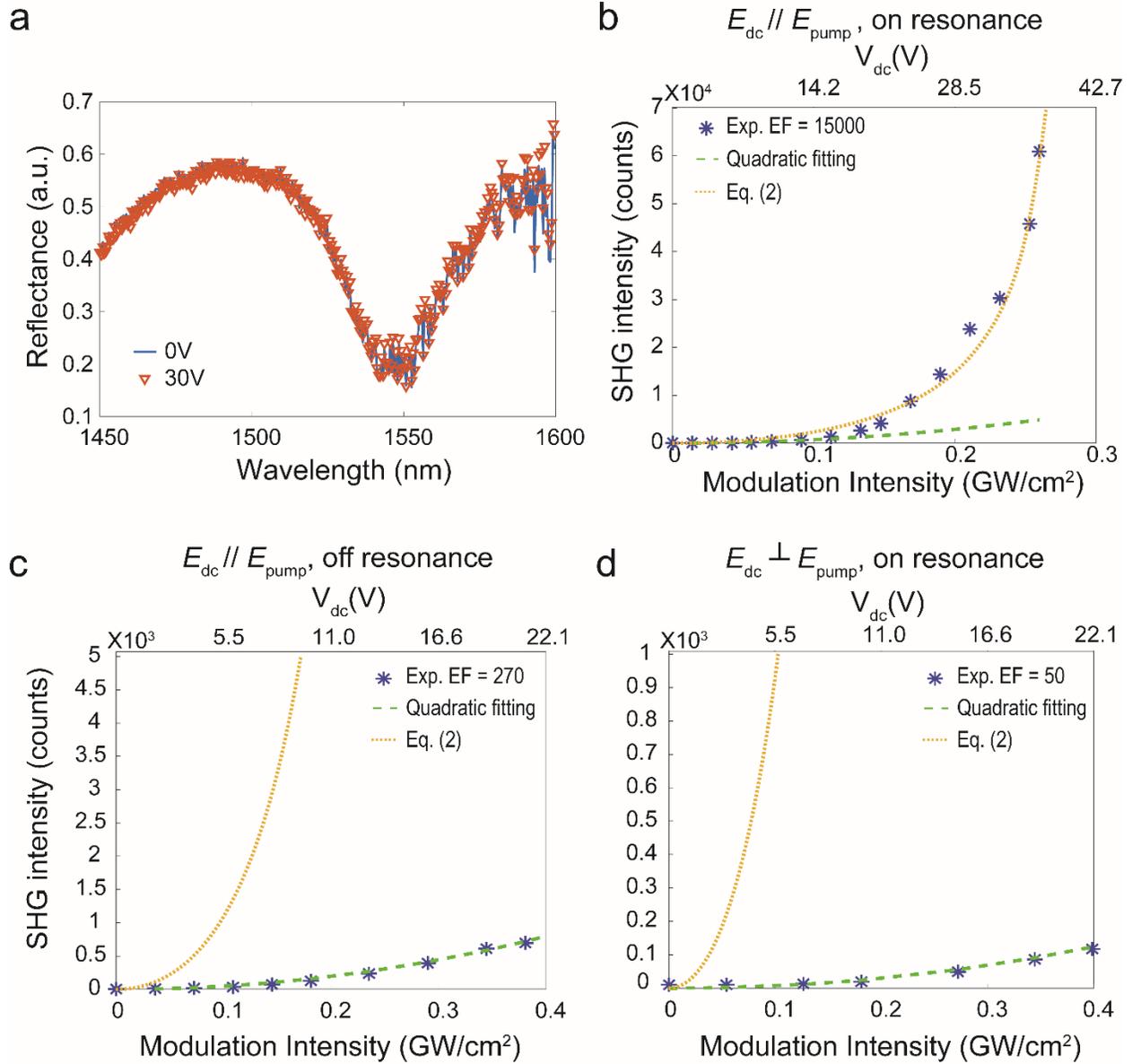

**Fig. 3. Experimental demonstration of parametrically enhanced SHG.** (**a**) Measured linear reflectance of fundamental optical field at dc voltage of 0V and 30 V shows the same resonance around 1550 nm, which excludes the possibility that dc field alone can change the resonance property of the nanoantennas. (**b**) – (**d**) dc electrical field tunable SHG. The modulated intensity is related to dc electric field and optical field, and the corresponding dc voltage ($V_{dc}$) is labelled on the upper axis. The blue asterisks, green dashed line and yellow dotted line represent the measured SHG intensity, the quadratic fitting of the measured signals and the fitting using equation (2) with the simulated $\Delta\varphi$ extracted from Fig. 2c, respectively. When the incident optical field (150 µW, 1530nm) hits the system's resonance and aligns with the dc electric field, strong dynamic

modulation occurs (**b**). The measured SHG signal fits well with equation (2) and achieves an enhancement factor of 15000. However, when the frequency of fundamental optical field (1 mW, 1400 nm) is off resonance, a quadratic SHG dependence on dc electric field is observed with a small enhancement factor of 270. In addition, as a control experiment, we rotated the antennas array by 90 degrees so that dc electric field is perpendicular to the optical field while maintaining the same resonance enhancement effect. In this condition (**c**), even though the fundamental optical field is on resonance, we cannot obtain a super-quadratic dc electric field dependence, due to a weak dynamic modulation ($\chi^{(3)}_{yxxy} < \chi^{(3)}_{yyyy}$).

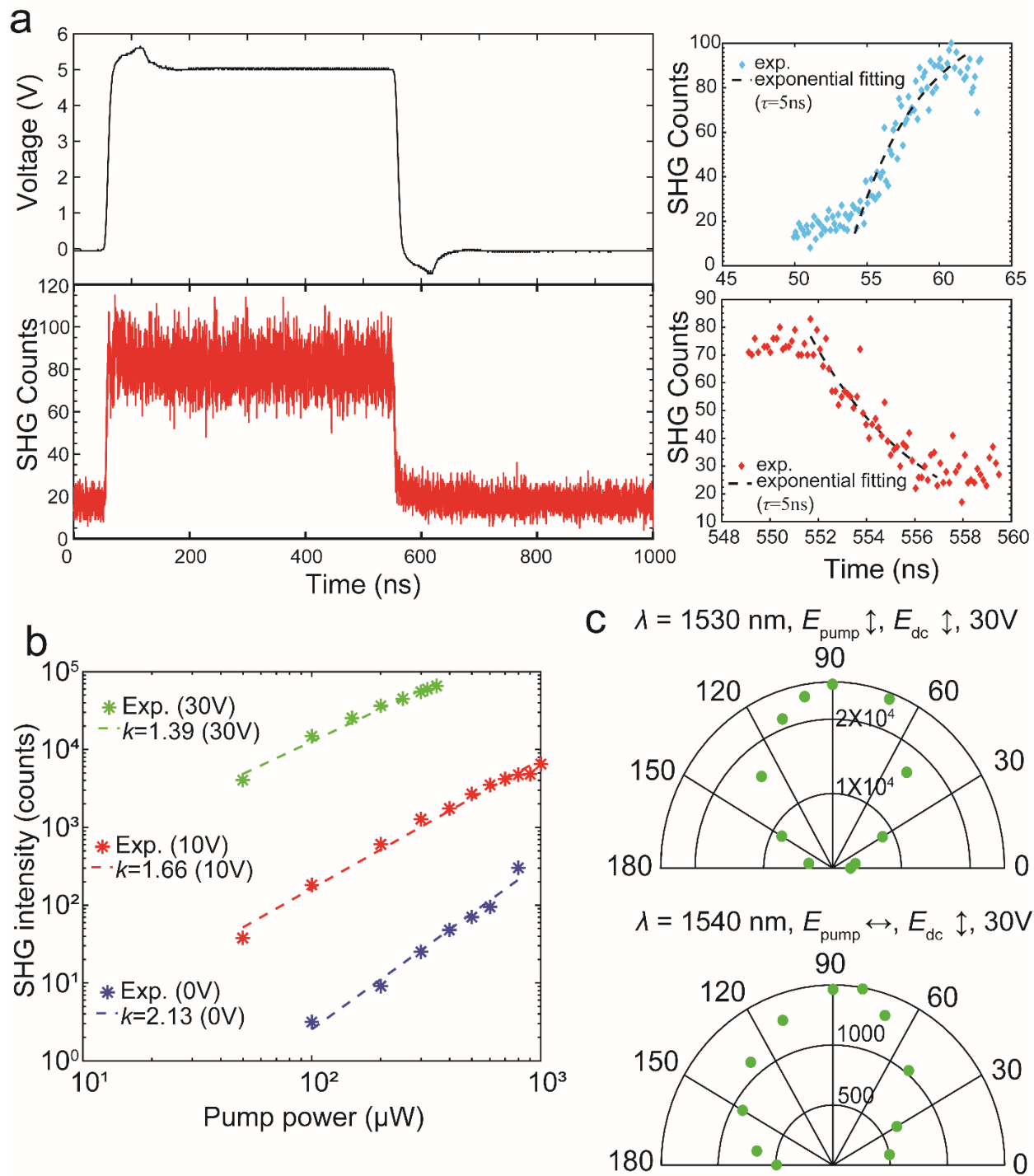

**Fig. 4** (**a**) Time-resolved trace of SHG (bottom panel) modulated by a square-wave electrical signal (top panel) at 1 MHz repetition rate. The SHG signal precisely follows the shape of the voltage pulses, indicating a 5 ns response time (extracted from the fittings of the rising and falling edges of SHG signal) only limited by RC constant of the device. (**b**) The SHG signal as a function of the power of the fundamental optical pump light with increased dc voltages. The dashed lines are least-square fits to experimental data (stars). As the dc voltage increases, the measured SHG signal shows a decrease in the power dependence on the fundamental intensity. This is due to the carrier screening effect of the built-in electric field. (**c**) Polarization properties of SHG signals under different conditions. The output SHG polarization dependence with an on-resonance incidence ($\lambda_{pump}$ =1530nm) polarized along the dc electric field (top), and the output SHG polarization dependence with an on-resonance incidence ($\lambda_{pump}$ =1540nm) polarized perpendicular to the dc electric field. With dynamic modulation (top), SHG signal is much stronger than that shown in the bottom polar plot. The polarization of SHG is dominated by the dc electric field no matter what polarization state of the optical pump field is, which agrees with the theory.